\documentclass[traditabstract]{aa}
\pdfoutput=1
\usepackage[pdftex]{graphicx}
\usepackage{natbib,amssymb,amsmath,url}
\usepackage{hyperref}
\usepackage{color}

\newcommand{\plotwd}{9.cm}
\newcommand{\plotwdtwo}{18.cm}

\begin{document}
\title{Probing large-scale structure with large samples of X-ray selected AGN: I Baryonic acoustic oscillations}
\titlerunning{Probing BAO with X-ray selected AGN}
\author{Gert H\"utsi \inst{1,2}, Marat Gilfanov \inst{1,3,4}, Alexander Kolodzig \inst{1}, Rashid Sunyaev \inst{1,3}}
\institute{Max-Planck-Institut f\"ur Astrophysik, Karl-Schwarzschild-Str. 1, 85741 Garching, Germany \\ \email{gert@mpa-garching.mpg.de} \and Tartu Observatory, T\~oravere 61602, Estonia \and Space Research Institute of Russian Academy of Sciences, Profsoyuznaya 84/32, 117997 Moscow, Russia \and Kazan Federal University, Kremlevskaya str. 18, 420008, Kazan, Russia}
\date{Received / Accepted}

\abstract{We investigate the potential of large X-ray-selected AGN samples for detecting baryonic acoustic oscillations (BAO).  Though AGN selection in X-ray band is very clean and efficient, it does not provide redshift information, and thus needs to be complemented with an optical follow-up. The main focus of this study is (i) to find the requirements needed for the quality of the optical follow-up and (ii) to formulate the optimal strategy of the X-ray survey,  in order to detect the BAO. We demonstrate that redshift accuracy of $\sigma_0=10^{-2}$ at $z=1$ and the catastrophic failure rate of $f_{\rm fail}\la 30\%$ are sufficient for a reliable  detection of  BAO in future X-ray surveys. Spectroscopic quality redshifts ($\sigma_0=10^{-3}$ and  $f_{\rm fail}\sim 0$) will boost  the confidence level of the BAO detection by a factor of $\sim 2$.  
For meaningful detection of BAO, X-ray surveys of moderate depth of $F_{\rm lim}\sim {\rm few~} 10^{-15}$ erg/s/cm$^2$ covering sky area from a few hundred to  $\sim$ten thousand square degrees are required. The optimal strategy for the BAO detection does not necessarily require full sky coverage. For example, in a 1000-day long survey by an eROSITA type telescope, an optimal strategy would be to survey a sky area of  $\sim 9000$ deg$^2$, yielding a $\sim 16\sigma$ BAO detection. A similar detection will be achieved by ATHENA+ or WFXT class telescopes in a survey with a duration of 100 days, covering a similar sky area. XMM-Newton can achieve a marginal BAO detection in a 100-day survey covering $\sim 400$ deg$^2$. These surveys would demand a moderate-to-high cost in terms the optical follow-ups, requiring determination of redshifts of $\sim 10^5$ (XMM-Newton) to $\sim 3\times 10^6$ objects (eROSITA, ATHENA+, and WFXT) in these sky areas.}
\maketitle

\section{Introduction}
Mapping the large-scale structure (LSS) in the low redshift Universe with modern redshift surveys, together with cosmic microwave background (CMB) \citep{2000Natur.404..955D,2000ApJ...545L...5H,2013ApJS..208...19H,2013arXiv1303.5076P} and type Ia supernova \citep{1998AJ....116.1009R,1999ApJ...517..565P} measurements, has helped to establish the current standard cosmological model -- the $\Lambda$CDM model. The initial discovery of the baryonic acoustic oscillations (BAO) \citep{1970Ap&SS...7....3S,1970ApJ...162..815P} in the two-point clustering statistics of the blue-band 2dFGRS\footnote{\url{http://www2.aao.gov.au/2dfgrs/}} and r-band-selected SDSS\footnote{\url{http://www.sdss.org/}} luminous red galaxy (LRG) samples (redshifts $z\sim 0.2$ and $z\sim 0.35$, respectively) \citep{2005MNRAS.362..505C,2005ApJ...633..560E} has initiated significant efforts to try to measure this signal, which provides a theoretically well-understood standard ruler, at ever higher redshifts, this way mapping out the low redshift expansion history of the Universe in great detail.\footnote{For a recent review see \citet{2013PhR...530...87W}.} 

Detecting BAO demands major observational efforts because this relatively weak signal, $\sim 5\%$ modulation in the matter power spectrum, can only be detected by combining large survey volumes with high enough sampling density, to overcome cosmic variance and shot noise, respectively. This leads to the typical number of redshifts obtained in these surveys ranging from $\sim 10^5$ up to $\sim 10^6$. The measurements of BAO in the galaxy two-point function have been extended to redshifts of $z\sim 0.55$ and $z\sim 0.7$ by the BOSS\footnote{\url{http://www.sdss3.org/surveys/boss.php}} and WiggleZ\footnote{\url{http://wigglez.swin.edu.au/‎}} surveys, using LRGs and emission line galaxies (ELGs), respectively \citep{2012MNRAS.427.3435A,2011MNRAS.418.1707B}. By exploiting the correlations in quasar Lyman-{$\alpha$} forest fluctuations, the BAO detection has been pushed up to $z\sim 2.3$ \citep{2013A&A...552A..96B,2013JCAP...04..026S}. Even though at the moment there are no BAO detections in the redshift range $z\sim 1-2$, it will be covered by the upcoming eBOSS\footnote{\url{http://www.sdss3.org/future/eboss.php}} survey. The measurement at even higher redshifts, $z\sim 3$, will be achieved by the HETDEX\footnote{\url{http://hetdex.org/}} survey. 

The eBOSS will use ELGs up to redshifts $z\sim 1$ and beyond that $\sim 750\, 000$ quasars (QSOs) will be used to sample cosmic density field over $\sim 7500$ deg$^2$, i.e., with a sampling rate of $\sim 100$ deg$^{-2}$. This is a factor of $\sim 2$ higher than a predicted number density of the X-ray active galactic nuclei (AGN) with $z>1$ from the upcoming eROSITA\footnote{\url{http://www.mpe.mpg.de/erosita/}} \footnote{\url{http://hea.iki.rssi.ru/SRG/}} all-sky-survey (eRASS) \citep[e.g.,][]{2012arXiv1209.3114M,2013A&A...558A..89K}.

Since the emission of X-rays is a generic feature accompanying AGN activity, the AGN selection in X-ray band is very effective \citetext{e.g., \citealp{2005ARA&A..43..827B}} and certainly much cleaner than the selection in other, lower energy wavebands. Since X-rays do not penetrate Earth's atmosphere, these observations have to be carried out in space, which severely limits the size of the achievable collective area. Despite that, as we know for example from the Chandra\footnote{\url{http://chandra.harvard.edu/‎}} deep field measurements \citep{2003AJ....126..539A,2011ApJS..195...10X}, the amazing AGN densities of $\sim 10^4$ deg$^{-2}$ should be achievable. However, to sample efficiently cosmic LSS, such high densities are completely unnecessary, and even modest AGN number densities, as achievable with eRASS in combination with large survey volumes, competitive measurements of two-point clustering statistics are possible.

It is clear that to obtain distance information, X-ray surveys have to be complemented with optical spectroscopic follow-up. This is exactly analogous to the optical redshift surveys, only the imaging part is replaced with imaging in the X-ray band.

In this paper we investigate the potential of the X-ray-selected AGN for probing the cosmic LSS in detail. For a clear benchmark of achievable quality, we
focus on the ability to detect the BAO in the clustering power spectrum. Since the BAO represent a $\sim 5\%$ modulation on top of the smooth broad-band spectral component, the smooth part itself can be detected with an order-of-magnitude higher signal-to-noise ratio.

One might wonder why attempt to use X-ray AGN for measuring the BAO. After all, almost all that matters for getting a BAO measurement is a large, uniformly covered survey volume combined with a large number of measured redshifts. Beyond the primary requirement for the redshifts to be obtained easily, the particular type of object used is only of secondary importance. Indeed, this is the way most of the upcoming BAO surveys are optimized; for example, for ELGs one only needs to detect the location of a few emission lines without needing to go down to the continuum level. 

In addition to a high enough sampling density of the LSS tracer objects, a  second important factor is their clustering strength. Indeed, because the signal-to-noise per Fourier mode scales as a product of the power spectrum amplitude and comoving number density, an increase in the clustering bias by a factor of two leads to the same signal-to-noise even if the sampling density is reduced by a factor of four. This in fact is an advantage of the X-ray-selected AGN, since they are more strongly clustered than optically selected QSOs or ELGs. Also, as it turns out, X-ray-selected AGN are numerous enough to efficiently probe LSS at redshifts $z\sim 1$.

In addition, it would be helpful if the cosmological BAO surveys could also be used for studying the astrophysics of the target objects. And this, we argue, is surely the case with X-ray-selected AGN. Indeed, the clean sample of AGN detected in X-ray band would certainly facilitate the study of accreting supermassive black holes (SMBHs) in the centers of galaxies, arguably one of the most remarkable discoveries of modern astrophysics. Also, additional synergetic effect might be expected from the fact that imaging and necessary spectroscopic follow-up are done in completely separate parts of the electromagnetic spectrum, this way helping to probe the physics of the AGN in a somewhat broader context. 
It is also important to realize that, no matter what, the prominence of the topic of AGN/SMBH evolution means that the optical follow-up of X-ray AGN samples detected by upcoming surveys like eROSITA \citep{2010SPIE.7732E..23P} will be done in one way or the other, and so using these AGN samples as probes of the LSS can at least be considered as an auxiliary research topic.

One might think that, in contrast to cosmology, detailed AGN studies do not possibly need to follow up such a large number ($\sim 10^6$) of objects. But if one wishes to measure AGN clustering with any reasonable (e.g., $\sim 10\%$) accuracy in several luminosity and redshift bins, and maybe also slice the data according to some other measurables, one cannot do with much smaller sample sizes. Our ability to constrain galaxy evolution models has benefited enormously from the availability of huge numbers of spectra from the LSS surveys, often driven mostly by cosmological needs. Similar gains should also be expected for the AGN science.

Owing to the importance of the optical follow-up for turning the X-ray selected AGN samples into genuine LSS probes, in this paper we aim to derive the necessary criteria for the quality of the spectroscopic/photometric redshifts. As stated above, we focus on the ability to detect the BAO; that is to say, the corresponding broad-band clustering signal is then detectable with an order of magnitude higher signal-to-noise.

Our paper is organized as follows. In Section~\ref{sec2} we present an initial feasibility study, Section~\ref{sec3} provides a short description of the modeling details, followed by our main results in Section~\ref{sec4}. Our summary and conclusions are given in Section~\ref{sec5}.  

Throughout this paper we assume flat $\Lambda$CDM cosmology with $\Omega_m=0.3$, $\Omega_b=0.05$, $h=0.7$, and $\sigma_8=0.8$.

\section{Initial feasibility study}\label{sec2}

\begin{figure}
\centering
\includegraphics[width=\plotwd]{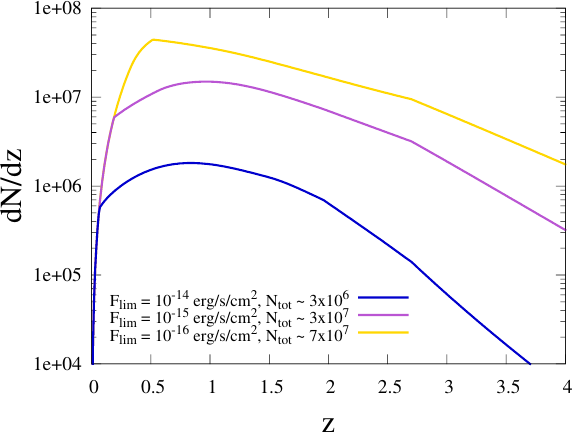}
\caption{Redshift distributions of X-ray-selected AGN, assuming the whole sky coverage and soft-band limiting fluxes as shown in the legend. Total numbers of AGN are also given there.}
\label{fig1}
\end{figure}

To assess the potential usefulness of X-ray-selected AGN for probing LSS, we start here with a brief feasibility study. Because the X-ray instruments are typically most efficient in the soft X-ray band, we focus on X-ray AGN selection in the $0.5-2$ keV energy range. To calculate the AGN redshift distributions for various limiting X-ray fluxes $F_{\rm lim}$, we use the soft-band AGN luminosity function as determined by \citet{2005A&A...441..417H}, along with corrections for $z>2.7$ as proposed in \citet{2009ApJ...693....8B} (see also \citet{2013A&A...558A..89K}). The AGN luminosity function of \citet{2005A&A...441..417H} only includes Type I AGN. Therefore it somewhat underpredicts the total number of objects, while the discrepancy with observed source counts increases toward the low flux end. As long as  the correlation properties of Type I and Type II AGN are similar,  this is not critically important for the main purpose of our calculations, but it will lead to underestimating of the confidence level of the BAO detection, predicted below.

In Fig.~\ref{fig1} we show the resulting AGN redshift distributions ${\rm d}N/{\rm d}z$ for three $F_{\rm lim}$ values: $10^{-14}$, $10^{-15}$, and $10^{-16}$ erg/s/cm$^2$. Here the full sky coverage is assumed. The first of the above limiting flux values is typical of the upcoming eRASS \citetext{e.g., \citealp{2013A&A...558A..89K}}. The second, $10^{-15}$ erg/s/cm$^2$, is approximately the soft-band limiting point source flux for the existing $\sim 2$ deg$^2$ XMM-COSMOS field \citep{2007ApJS..172...29H} and, depending on the final survey strategy,  may also be almost reachable (albeit close to the confusion limit) in `pole regions' of the eRASS where over $\sim 100$ deg$^2$ the exposure will be significantly deeper than the sky average \citetext{e.g., \citealp{2013A&A...558A..89K}}. The smallest limiting fluxes we consider correspond to $10^{-16}$ erg/s/cm$^2$, which is about an order of magnitude more than is typical of the current deepest X-ray fields, i.e., Chandra Deep Field South \citep{2011ApJS..195...10X} and Chandra Deep Field North \citep{2003AJ....126..539A}.
 
In Fig.~\ref{fig1} we see that, irrespective of the $F_{\rm lim}$ value, the X-ray AGN distribution peaks around $z\sim 1$. The expected total number of AGN over the full sky computed with the AGN X-ray luminosity function of \citet{2005A&A...441..417H} is $\sim 3\times 10^6$, $\sim 3\times 10^7$, and $\sim 7\times 10^7$ for $F_{\rm lim}=10^{-14}$, $10^{-15}$, and $10^{-16}$, respectively. In the redshift range $z=0.5-2.5$, which is used extensively throughout the rest of the paper, the corresponding numbers reduce to $\sim 2\times 10^6$, $\sim 2\times 10^7$, and $\sim 5\times 10^7$, with corresponding sky densities of $\sim 60$, $\sim 500$, and $\sim 1300$ deg$^{-2}$, respectively. It is interesting to note that once the redshift interval $z=0.5-2.5$ is divided into three subintervals $z=0.5-1.0$, $z=1.0-1.5$, and $z=1.5-2.5$ then, quite independently of the value of $F_{\rm lim}$, each of these contains approximately one third of the objects.

To investigate whether these numbers are large enough to efficiently probe the LSS, we have to include some knowledge of the X-ray AGN clustering strength. Several recent studies have shown \citetext{e.g., \citealp{2011ApJ...736...99A,2012ApJ...746....1K,2013MNRAS.430..661M}} that the clustering of X-ray-selected AGN is accounted for well if they populate group-size dark matter (DM) halos. Throughout this paper we assume a single effective host DM halo mass of $M_{\rm eff}=2\times 10^{13}$ M$_\odot$, which is compatible with clustering bias measurements, as given in \citet{2011ApJ...736...99A} up to redshift of $\sim 3$.

\begin{figure}
\centering
\includegraphics[width=\plotwd]{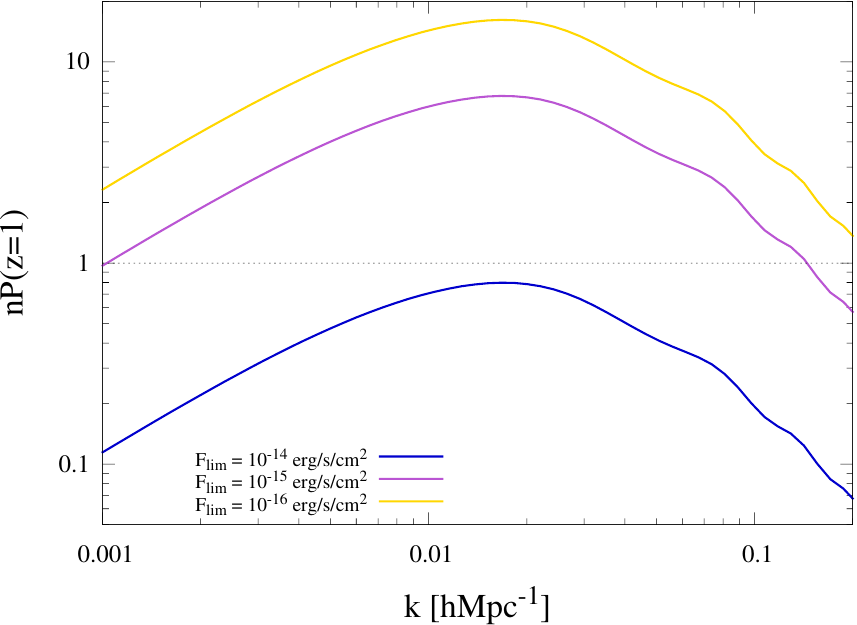}
\caption{Signal-to-noise per Fourier mode at redshift $z=1$ for flux-limited samples of X-ray AGN. The limiting flux values are shown in the legend. The linear clustering bias is assumed to correspond to DM halos with mass $M_{\rm eff}=2\times 10^{13}$ M$_\odot$.}
\label{fig2}
\end{figure}

In Fig.~\ref{fig2} we show the achievable signal-to-noise per Fourier mode at redshift $z=1$ for the same limiting fluxes as in Fig.~\ref{fig1}. To calculate the matter power spectrum, we use the approximate fitting forms as given in \citet{1998ApJ...496..605E}. The clustering bias parameter is taken from \citet{2001MNRAS.323....1S} for the effective DM halo mass of $2\times 10^{13}$ M$_{\odot}$, as stated above. The discreteness noise due to finite sampling density is given as usual as $1/n$, where $n$ is the comoving number density of AGN. We see that for limiting fluxes of $10^{-15}$ and $10^{-16}$ erg/s/cm$^2$, the achievable signal-to-noise is significantly higher than one for a broad range of wavenumbers. Shown in Fig.~\ref{fig2} is the signal-to-noise for a Fourier mode, i.e. for the unbinned power spectra. In fact, one does not need such high sampling density,\footnote{In practice, for two-point clustering measurements, an effective signal-to-noise ratio of $nP\sim 3$ is often more than enough, with significantly higher values leading to a wasteful oversampling of the density field.} and in terms of optimizing the observational strategy, it would be wiser to move to a new field, instead of integrating too long at the same position to achieve lower $F_{\rm lim}$.

With eRASS type of sensitivities, i.e. $F_{\rm lim}\simeq 10^{-14}$, one typically undersamples the large-scale density field. Even then, the above mild undersampling can be hugely compensated by a large survey volume, which in the end can still lead to a tight measurement of the two-point clustering signal \citep{2013A&A...558A..90K}.

Thus, we conclude that the with the limiting flux in the range $\sim 10^{-16}-10^{-14}$ erg/s/cm$^2$, which is typical for modern narrow and large field surveys, there is enough X-ray AGN to efficiently probe the LSS at redshifts $z\sim 0-3$, with the tightest clustering measurements possible around redshift $z\sim 1$. 

\section{Modeling details}\label{sec3}

Compared to our earlier study in \citet{2013A&A...558A..90K}, we have implemented several improvements in our two-point function calculations: (i) in addition to the auto-spectra we include the information encoded in cross-spectra, (ii) we include linear redshift-space distortions, (iii) BAO damping is treated by using the resummed Lagrangian perturbation theory (LPT) approach by \citet{2008PhRvD..78h3519M}. Most important, since the main focus of this study is to try to determine the criteria necessary for the quality of the optical follow-up, we have a significantly more elaborate treatment for the photometric redshift (photo-z) errors.\footnote{In \citet{2013A&A...558A..90K}, to approximate the impact of photo-z errors, we simply adjusted the effective width of the redshift bins accordingly.}

To flexibly treat for the smoothing caused by the photometric redshifts, we divide the full survey volume into narrow photo-z bins and calculate projected power spectra within bins, i.e. auto-spectra, and also cross-spectra between the bins. Even if full spectroscopic redshifts are available, it might be beneficial to carry out clustering analysis by dividing survey volume into narrow redshift bins and calculate projected auto and cross power spectra. From these spectra a full three-dimensional power spectrum for the whole volume can be recovered up to some wavenumber $k_{\max}$ that depends on how narrow redshift binning one has decided to use. This contrasts with how spectroscopic galaxy surveys are currently analyzed. There one typically makes a single three-dimensional power spectrum measurement using the whole survey volume. But this way one loses sensitivity to evolutionary trends of the clustering properties with redshift. Also, by calculating projected auto- and cross-spectra in narrow bins, one does not need to assume any fiducial cosmological model in order to convert redshifts to comoving distances, and thus data analysis can be done once and for all. Even though the consequent model fitting is computationally more demanding, especially when the number of redshift bins rises significantly, e.g. $\sim 100$, this is not really an obstacle for modern computational technology.

\subsection{Photometric redshifts}
We assume that the conditional probability distribution for photometric redshift $z_p$ given spectroscopic redshift $z$ can be decomposed as
\begin{equation}\label{eq01}
P(z_p|z)=(1-f_{\rm fail})P_1(z_p|z) + f_{\rm fail}P_2(z_p|z)\,.
\end{equation}
Here $P_1$ is assumed to be a truncated Gaussian with the requirement that $P_1(z_p<0|z)=0$, and thus the probability distribution function (pdf) with appropriate normalization is given as
\begin{equation}
P_1(z_p|z)=\sqrt{\frac{2}{\pi}}\frac{\exp\left[-\frac{(z-z_p)^2}{2\sigma^2(z)}\right]}{\sigma(z)\left[1+{\rm erf}\left(\frac{z}{\sqrt{2}\sigma(z)}\right)\right]}\,.
\end{equation}
For $\sigma(z)$ we assume a commonly used form
\begin{equation}\label{eq03}
\sigma(z)=\sigma_0(1+z)\,.
\end{equation}
The quantity $f_{\rm fail}$ in Eq.~(\ref{eq01}) represents the fraction of catastrophic errors. For simplicity we assume that in the case of a  catastrophic failure, the measured redshift value can be represented by a random number uniformly distributed between $0$ and some $z_p^{\rm max}$, i.e.,
\begin{equation}
P_2(z_p|z)=\frac{1}{z_p^{\rm max}}\,.
\end{equation}
As a default value we use $z_p^{\rm max}=5$ throughout this paper.\footnote{As it turns out, our results have little dependence on $z_p^{\rm max}$, e.g., by changing its value by $\pm 2$ our results remain almost the same.}

The above simplistic model for photometric redshifts is relatively conservative, since in more realistic cases, catastrophic failures might have a more complex distribution, e.g., might form a secondary bump in redshift distribution. If this (somewhat more structured) distribution is known, it can be used, in principle, to enhance the confidence level of the BAO detection (compared to the case with broad uniform catastrophic error distribution).

Thus, we have assumed photo-z errors to be independent of luminosity and the fraction of catastrophic errors to be independent of redshift. In realistic situation this is certainly not true (see \citet{2011ApJ...742...61S} for current performance of photometric redshifts for X-ray-selected AGN), but since in our study the AGN clustering is taken to be independent of luminosity and we consider only a cumulative clustering signal over broad redshift range, one can always use effective luminosity- and redshift-averaged values for $\sigma_0$ and $f_{\rm fail}$.

The pdf of true redshifts corresponding to the objects selected in the $i$-th photo-z bin $z_{p_1}^{(i)}-z_{p_2}^{(i)}$ is given as \citetext{see also, e.g., \citealp{2003ApJ...595...59B}}
\begin{equation}
f^{(i)}(z)=f(z)\int_{z_{p_1}^{(i)}}^{z_{p_2}^{(i)}}P(z_p|z)\,{\rm d}z_p\,.
\end{equation}
Here $f(z)$ is the true underlying full redshift distribution function, i.e.
\begin{equation}
f(z)=\frac{\frac{{\rm d}N}{{\rm d}z}}{\int\frac{{\rm d}N}{{\rm d}z}\,{\rm dz}}=\frac{1}{N_{\rm tot}}\frac{{\rm d}N}{{\rm d}z}\,.
\end{equation}
Taking $P(z_p|z)$ from Eq.~(\ref{eq01}) we arrive at the result
\begin{align}\label{eq07}
f^{(i)}(z)&= f(z)\Biggl[(1-f_{\rm fail})\frac{{\rm erf\left(\frac{z_{p_2}^{(i)}-z}{\sqrt{2}\sigma(z)}\right)}-{\rm erf\left(\frac{z_{p_1}^{(i)}-z}{\sqrt{2}\sigma(z)}\right)}}{1+{\rm erf}\left(\frac{z}{\sqrt{2}\sigma(z)}\right)}+\nonumber\\
&+ f_{\rm fail}\frac{z_{p_2}^{(i)}-z_{p_1}^{(i)}}{z_p^{\rm max}}\Biggr]\,,
\end{align}
with $\sigma(z)$ given by Eq.~(\ref{eq03}).

Thus, apart from the photo-z bin boundary values $z_{p_1}^{(i)}$, $z_{p_2}^{(i)}$, and underlying true redshift distribution $f(z)$ (see Fig.~\ref{fig1}), in our model the radial selection functions $f^{(i)}(z)$ are fully determined by two parameters only: $\sigma_0$ and $f_{\rm fail}$. In this paper we assume these parameters to stay within ranges $\sigma_0=10^{-3}-10^{-1}$ and $f_{\rm fail}=0-0.5$. It should be clear that for the large-scale clustering analysis the highest accuracy photo-z choice $\sigma_0=10^{-3}$ is in practice almost equivalent to the availability of the full spectroscopic measurements.

\begin{figure}
\centering
\includegraphics[width=\plotwd]{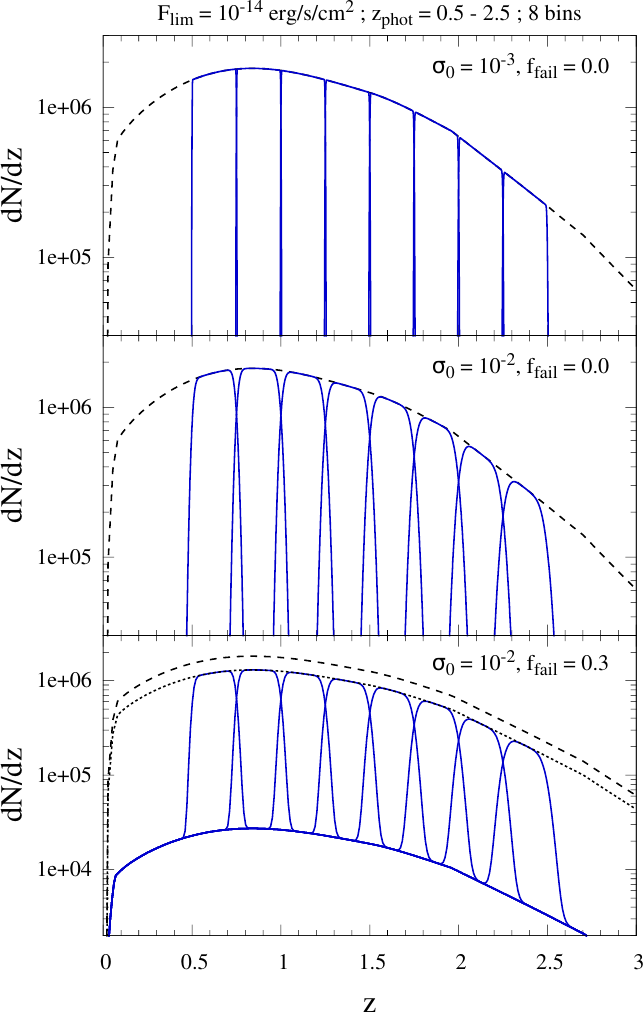}
\caption{Radial selection functions for eight equal size photo-z bins between $z_p=0.5-2.5$. Three different combinations of $\sigma_0$ and $f_{\rm fail}$ are used and the limiting flux is assumed to be $F_{\rm lim}=10^{-14}$ erg/s/cm$^2$. The dashed lines show the full underlying spectroscopic redshift distribution.}
\label{fig3}
\end{figure}

In Fig.~\ref{fig3} we show some examples of the resulting ${\rm d}N^{(i)}/{\rm d}z\equiv N_{\rm tot}f^{(i)}(z)$ distributions with solid lines. Here we assume eight equal size photo-z bins between $z_p=0.5-2.5$ and limiting flux $F_{\rm lim}=10^{-14}$ erg/s/cm$^2$. Three panels correspond to various choices of $\sigma_0$ and $f_{\rm fail}$ as indicated in the upper righthand corners. The dashed line corresponds to the full underlying ${\rm d}N/{\rm d}z$ distribution.

\subsection{Choice of the  redshift bins width}

How many photo-z bins should one use? In reality there is no need to use redshift bins that are much smaller than $\sigma(z)$, since photometric redshift errors in that case would highly correlate neighboring bins, and by using more of them we only make our calculations computationally more demanding, without gaining any extra information.

There is also another bound for the redshift bin size one has to consider. Even when photo-z becomes very accurate, say by approaching the accuracy of spectroscopic redshifts, we still should not use extremely narrow redshift bins, since in that case we would effectively incorporate nonlinear radial modes into our clustering analysis. To avoid these complications one typically applies some wavenumber cutoff to eliminate nonlinear Fourier modes.

\begin{figure}
\centering
\includegraphics[width=\plotwd]{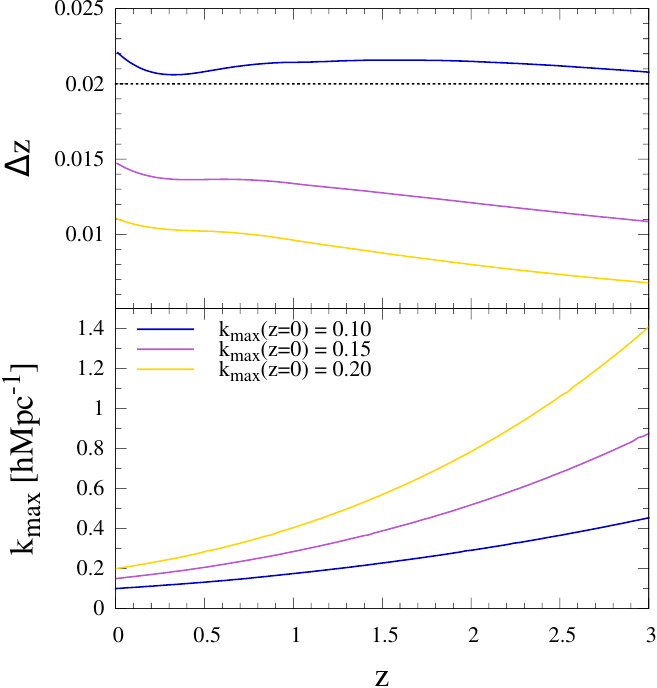}
\caption{Maximum wavenumber as a function of redshift, $k_{\max}(z)$, calculated from Eq.~(\ref{eq09}) for three different values of $k_{\max}(0)$ (lower panel) and corresponding redshift intervals (upper panel).}
\label{fig4}
\end{figure}

The variance of the underlying DM density field at redshift $z$ can be expressed as
\begin{equation}
\sigma_{\rm DM}^2(z)=\frac{g^2(z)}{2\pi^2}\int k^2P(k|z=0)\,{\rm d}k\,,
\end{equation}
where $g(z)$ is the linear growth factor and $P(k|z=0)$ the matter power spectrum at $z=0$. One typically cuts this integral at some $k_{\max}$, such that the resulting $\sigma^2_{\rm DM}(z)$ stays safely below one. At $z=0$ one usually assumes $k_{\max}$ is in the range $\sim 0.1-0.2$ $h\,$Mpc$^{-1}$, while at higher redshifts one is able to accommodate higher values for $k_{\max}$. Here we obtain $k_{\max}$ as a function of redshift by demanding that the above integral for $\sigma^2_{\rm DM}(z)$ with appropriately adjusted $k_{\max}(z)$ stays the same as in $z=0$ case; i.e.,
\begin{equation}\label{eq09}
g^2(z)\int_0^{k_{\max}(z)}\mkern-18mu k^2P(k|z=0){\rm d}k=g^2(0)\int_0^{k_{\max}(0)}\mkern-18mu k^2P(k|z=0){\rm d}k.
\end{equation}
The numerical solution $k_{\max}(z)$ for the above equation for various values of $k_{\max}(0)$ are shown in the lower panel of Fig.~\ref{fig4}.

The maximum wavenumber $k_{\max}(z)$ corresponds to the minimal `allowed' comoving spatial scale $r_{\min}(z)=2\pi/k_{\max}(z)$. When this spatial interval is placed at redshift $z$ and oriented along the line of sight, it corresponds to the redshift interval $\Delta z(z)$ as plotted in the upper panel of Fig.~\ref{fig4}. We see that for the conservative choice of $k_{\max}(0)=0.1$ $h\,$Mpc$^{-1}$, the redshift interval $\Delta z$ stays remarkably constant with a value close to $0.02$. In the following we adopt $\Delta z=0.02$ as the minimum allowed size for the redshift bin. 

As noted above, due to photo-z smoothing along the radial direction, there is no benefit of considering redshift bins significantly smaller than $\sigma(z)$. It turns out that by choosing $\Delta z=\max(\sigma_0,0.02)$ one already obtains well-converged results.

To fully eliminate small-scale nonlinear modes, one also has to impose some cutoff $\ell_{\max}$ for the angular multipole number. Throughout this paper we have used $\ell_{\max}=500$. The justification is the following. The AGN distribution peaks at $z\sim 1$, which corresponds to a comoving distance of $\sim 2300$ $h^{-1}\,$Mpc. The cutoff $k_{\max}(0)=0.1$ $h\,$Mpc$^{-1}$, which was also applied for the radial modes, corresponds to $\sim 0.2$ $h\,$Mpc$^{-1}$ at $z\sim 1$ (see Fig.~\ref{fig4}), which results in $\ell_{\max}\sim 0.2\times 2300\sim 500$. For simplicity, because we look at cumulative signal from broad redshift range ($z=0.5-2.5$) in this study, we keep this value fixed, independent of redshift.

\subsection{Auto- and cross-spectra}
In the linear regime the angular clustering spectra can be given as \citep{1994MNRAS.266..219F,2001ApJ...555..547H,2007MNRAS.378..852P,2012MNRAS.427.1891A}
\begin{equation}\label{eq10}
C_{\ell}^{(ij)}=\frac{2}{\pi}\int W_{\ell}^{(i)}(k)W_{\ell}^{(j)}(k)P(k|z=0)k^2\,{\rm d}k\,.
\end{equation}
If $i=j$ we obtain auto-spectra, and $i\ne j$ gives us cross-spectra between different redshift bins. The projection kernel for the $i$-th redshift bin, $W_{\ell}^{(i)}(k)$, is given as a sum of two parts
\begin{equation}
W_{\ell}^{(i)}(k)=W_{\ell}^{(i),1}(k)+W_{\ell}^{(i),2}(k)\,.
\end{equation}
Here the first part takes care of the projection in real space, while the second one includes the effect of redshift-space distortions. They are given as
\begin{align}
W_{\ell}^{(i),1}(k)&=\displaystyle\int j_{\ell}(kr)f^{(i)}(r)g(r)b(r)\,{\rm d}r\,,\\
W_{\ell}^{(i),2}(k)&=\displaystyle\int\Biggl[\frac{2\ell^2+2\ell-1}{(2\ell+3)(2\ell-1)}j_{\ell}(kr)-\nonumber\\
&-\frac{\ell(\ell-1)}{(2\ell-1)(2\ell+1)}j_{\ell-2}(kr)-\nonumber\\
&-\frac{(\ell+1)(\ell+2)}{(2\ell+1)(2\ell+3)}j_{\ell+2}(kr)\Biggr]f^{(i)}(r)g(r){\rm f}(r)\,{\rm d}r\,.\label{eq13}
\end{align}
Here the integrals are over comoving distance $r$, $j_{\ell}$ is the $\ell$-th order spherical Bessel function; $f^{(i)}(r)$ is the normalized distribution of comoving distances of LSS tracers (AGN) in the $i$-th photo-z bin, i.e., $f^{(i)}(r)=f^{(i)}(z)\frac{{\rm d}z}{{\rm d}r}$ with $f^{(i)}(z)$ given in Eq.~(\ref{eq07}); $g$ and $b$ are the linear growth factor and clustering bias parameter, respectively. In Eq.~(\ref{eq13}) ${\rm f}(r)\equiv\frac{{\rm d}\ln g}{{\rm d}\ln a}(r)$, with $a$ the scale factor, is the linear growth rate at comoving distance $r$. 

\subsection{BAO damping}
Since the focus of the AGN clustering analysis performed in this paper is the ability to detect the BAO, to be more accurate we have to include the damping of linear BAO due to nonlinear evolution of the cosmic density field. Here we adopt the two-point function results for DM halos derived from the resummed LPT by \citet{2008PhRvD..78h3519M}. Even with BAO damping included, we still use the simple form of Eq.~(\ref{eq10}) where the $k$-independent linear growth factor separates out. Namely, we make the following approximate replacement in Eq.~(\ref{eq10})
\begin{equation}
P(k|z=0)\longrightarrow\sqrt{\frac{P^{\rm LPT}(k,z_i)P^{\rm LPT}(k,z_j)}{g(z_i)g(z_j)}}\,,
\end{equation}
where $P^{\rm LPT}$ is the power spectrum calculated with LPT, and $z_i$ and $z_j$ are central redshifts in the $i$-th and $j$-th bins, respectively.

\begin{figure}
\centering
\includegraphics[width=\plotwd]{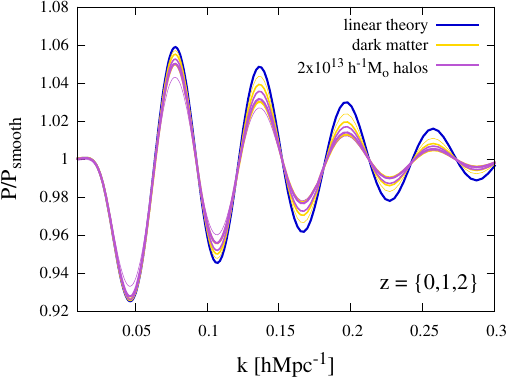}
\caption{Damping of BAO at redshifts $z=0,1,2$ for DM and for halos with mass $2\times 10^{13}$ M$_\odot$ compared to the linear BAO. Thinner lines correspond to higher redshifts.}
\label{fig5}
\end{figure}

In Fig.~\ref{fig5} we illustrate the damping of BAO at various redshifts ($z=0,1,2$) for DM halos with mass $2\times 10^{13}$ M$_\odot$ and similarly for the full DM distribution. Linear theory BAO is also shown for comparison. Thinner lines correspond to higher redshifts. As expected, for the DM the damping behavior as a function of redshift is simply monotonic, with the BAO steadily approaching linear theory prediction at higher redshifts. The low redshift behavior for the DM halos mimics the trend seen in the full DM distribution, albeit with somewhat reduced amplitude, but the evolutionary behavior has reversed at higher redshifts, with the BAO amplitude becoming more suppressed. This is because halos with masses $2\times 10^{13}$ M$_\odot$ at $z\sim 2$ are significantly rarer than similar mass halos at lower redshifts, and so in relative terms involve a significantly higher level of `nonlinear processing'. 

\subsection{Covariance matrix. Confidence level for BAO detection}
From the CMB measurements we know that the initial density fluctuations can be described as a Gaussian random field with very good accuracy \citep{2013arXiv1303.5084P}. Since the linear evolution does not change this statistical property, one can write (using Wick's theorem) the covariance matrix for the angular power spectra (which contain the full information content of the Gaussian field) as
\begin{eqnarray}
{\cal C}_{\ell}&\equiv&{\rm cov}\left(C_{\ell}^{(ij)}C_{\ell}^{(mn)}\right)=\nonumber\\
&=&\frac{1}{(2\ell+1)f_{\rm sky}}\left[\widetilde{C}_{\ell}^{(im)}\widetilde{C}_{\ell}^{(jn)}+\widetilde{C}_{\ell}^{(in)}\widetilde{C}_{\ell}^{(jm)}\right]\,;\label{eq15}
\end{eqnarray}
i.e., it can be factorized as a product of power spectra \citetext{e.g., \citealp{1994ApJ...426...23F,1997ApJ...480...22T,1999MNRAS.308.1179M,1999ApJ...527....1S,2009MNRAS.400..851S,2012MNRAS.427.1891A}}. Here the number of modes with multipole number $\ell$ for the full sky, i.e. $2\ell+1$, is reduced by a factor $f_{\rm sky}$, which is the fraction of sky covered by the survey. However, there is also a slight complication due to discrete sampling of the density field. Namely, the auto-spectra that enter the product in Eq.~(\ref{eq15}) get extra Poisson noise contribution; i.e.,
\begin{equation}
\widetilde{C}_{\ell}^{(ij)}\equiv C_{\ell}^{(ij)}+\frac{1}{{\cal N}^{(i)}}\delta_{ij}\,.
\end{equation}
Here $\delta_{ij}$ is the Kronecker delta and ${\cal N}^{(i)}$ the number of objects per steradian in the $i$-th redshift bin. For an in-depth discussion on covariance of auto- vs cross-spectra, see \citet{2009MNRAS.400..851S}. 
 
Equation~(\ref{eq15}) assumes that covariance matrix for the angular power spectra can be written separately for each multipole $\ell$; that is, it assumes that different multipoles are not coupled, which is only strictly true for the unmasked Gaussian random field. However, for reasonably large survey areas, the resulting mode coupling is fairly tightly localized in multipole space, and Eq.~(\ref{eq15}) is a very good approximation in practice. With $N$ redshift bins, each matrix ${\cal C}_{\ell}$ has a dimension of $\frac{N}{2}(N+1)\times\frac{N}{2}(N+1)$.

Having specified the covariance matrix, we can proceed to ask the question how well the BAO can be detected. First we have to specify a smooth template spectrum without acoustic features with respect to what the BAO is measured. Instead of the standard `no wiggle' form of \citet{1998ApJ...496..605E} we use a slightly different form, which we believe behaves somewhat better. The details are presented in Appendix~\ref{appa}. In reality, owing to the applied wavenumber cuts, these small deviations in the `no wiggle' spectral form have a negligible impact on our results.

To estimate how accurately the BAO can be measured, we look at the parametric class of models with power spectra given as
\begin{equation}
P^{A_{\rm BAO}}(k)=(1-A_{\rm BAO})P^{\rm NW}(k)+A_{\rm BAO}P(k)\,,
\end{equation}
i.e., models interpolating between the `no wiggle' spectral form and the form with a full $\Lambda$CDM BAO. Thus $A_{\rm BAO}$ can take the values between zero and one, with the fiducial value $A_{\rm BAO}=1$.

The angular spectra can then be given similarly; i.e.,
\begin{equation}
C_{\ell}^{(ij),A_{\rm BAO}}=(1-A_{\rm BAO})C_{\ell}^{(ij),{\rm NW}}+A_{\rm BAO}C_{\ell}^{(ij)}\,.
\end{equation}
Now one can write down the Fisher information (for detailed presentation of Fisher information in Gaussian cases see \citet{1997ApJ...480...22T}) for the parameter $A_{\rm BAO}$ from the spectra at fixed multipole $\ell$
\begin{eqnarray}
F_{\ell}^{A_{\rm BAO}}&=&\left(\frac{\partial{\mathbf C}_{\ell}}{\partial A_{\rm BAO}}\right)^T{\cal C}_{\ell}^{-1}\left(\frac{\partial{\mathbf C}_{\ell}}{\partial A_{\rm BAO}}\right)=\nonumber\\
&=& \left({\mathbf C}_{\ell}-{\mathbf C}_{\ell}^{\rm NW}\right)^T{\cal C}_{\ell}^{-1}\left({\mathbf C}_{\ell}-{\mathbf C}_{\ell}^{\rm NW}\right)\,,
\end{eqnarray}
where with ${\mathbf C}_{\ell}$ (${\mathbf C}_{\ell}^{\rm NW}$) we have denoted the vector with components $C_{\ell}^{(ij)}$ ($C_{\ell}^{(ij),{\rm NW}}$), i.e., in case of $N$ redshift bins, an $\frac{N}{2}(N+1)$-dimensional vector. Since all the multipoles are assumed to be independent, the full Fisher information for $A_{\rm BAO}$ and its standard error are given as
\begin{eqnarray}
F^{A_{\rm BAO}}&=&\sum_{\ell}F_{\ell}^{A_{\rm BAO}}\,,\\
\sigma_{A_{\rm BAO}}&=&\frac{1}{\sqrt{F^{A_{\rm BAO}}}}\,.
\end{eqnarray}
Since our fiducial $A_{\rm BAO}=1$, the confidence level (CL) for the BAO detection can be expressed as\footnote{One could have arrived at this result also by performing model comparison through a likelihood ratio test. For the Gaussian case, if the dimensionalities of the model parameter spaces differ by a single unit, the resulting $\Delta \chi^2$ is distributed as a $\chi^2$ distribution with one degree of freedom, and thus one model is preferred over the other with ${\rm CL}=\sqrt{\Delta \chi^2}$, which is equivalent to the result in Eq.~(\ref{eq22}).}
\begin{equation}\label{eq22}
{\rm CL}=\frac{1}{\sigma_{A_{\rm BAO}}}=\sqrt{F^{A_{\rm BAO}}}\,.
\end{equation}

In case we have $N$ redshift bins, the number of spectra one has to calculate to find CL for the BAO detection as presented above is $\frac{N}{2}(N+1)$. For a survey from $z=0.5$ up to $z=2.5$ with our minimal $\Delta z=0.02$, i.e. $N=100$, this leads to a total of $5050$ spectra. However, it is clear that cross-spectra between far away bins are quite small in amplitude, and thus in practice can be neglected.

To speed up the computation, we calculate CL of Eq.~(\ref{eq22}) in an iterative manner. (i) We start with auto-spectra $C_{\ell}^{(ii)}$ only; i.e. the initial covariance matrix dimension is $N\times N$. (ii) In the next step we include cross-spectra between the neighboring bins, i.e. $C_{\ell}^{(i,i+1)}$. (iii) Step by step we continue including $C_{\ell}^{(i,i+2)}$, $C_{\ell}^{(i,i+3)}$, etc., each time calculating CL from Eq.~(\ref{eq22}). (iv) Once the value for CL does not change by more than $1\%$, we consider the obtained CL to be converged and stop our calculation.

For small $\sigma_0$, we have a rather narrow redshift binning, so nearby bins might get very tightly correlated, thus leading to a singular power spectrum covariance matrix. As explained above, we are applying the minimal redshift bin size of $\Delta z=0.02$. This limiting bin size, together with double-precision numerical resolution, should be fine enough to make the covariance matrix numerically invertible. However, in practice, to avoid possible issues with numerical instability, we always use the Moore-Penrose pseudoinverse instead of the usual matrix inverse, as calculated via a singular value decomposition.

As stated above, for treating the BAO damping, we use the resummed LPT, but for the power spectrum covariance we still assume a simple Gaussian random field assumption. This is fully justified since we are applying a rather conservative maximum wavenumber cutoff of $k_{\max}(0)=0.1$ $h\,$Mpc$^{-1}$. In principle, one could also try to use smaller scale modes, but then the Gaussianity assumption is no longer justified, and one should start considering nonlinear mode-mode couplings, which one might hope to treat again within the LPT framework. But this task is beyond the scope of this work.

Furthermore, even with the currently applied high wavenumber cutoff, in a real survey with a non-trivial survey mask, one should not expect the nearby Fourier modes to be independent. As usual, this effective loss of modes due to survey mask is treated in a very simple way, where for the covered sky fraction of $f_{\rm sky}$, one is left with $f_{\rm sky}(2\ell+1)$ independent modes for multipole $\ell$. In reality the validity of the above approximation depends quite significantly on a particular survey geometry, with the above treatment being sufficient only for the cases where the survey footprint is mostly continuous and extending roughly equally in both angular directions.

\begin{figure}
\centering
\includegraphics[width=\plotwd]{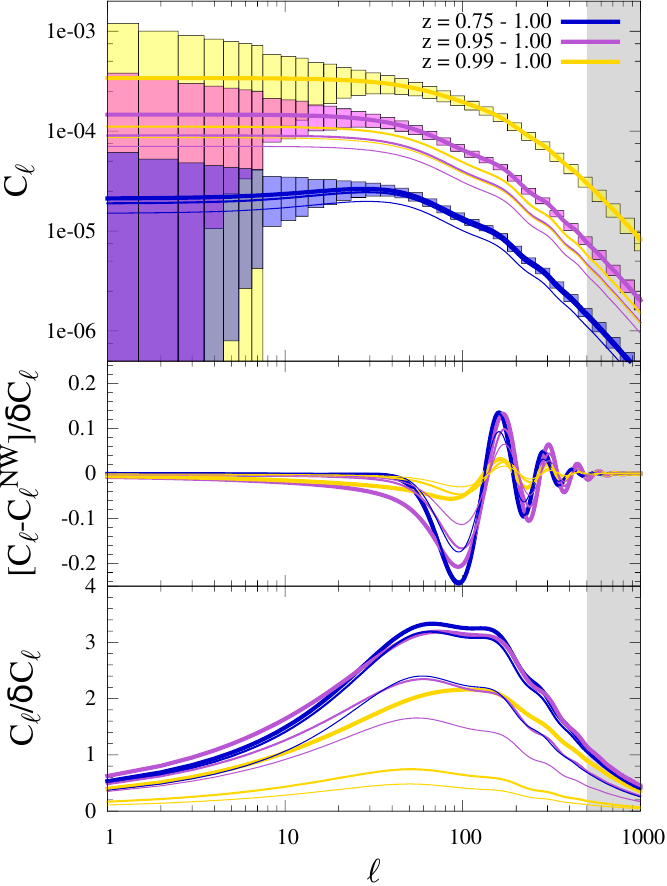}
\caption{Example angular auto-spectra (upper panel) and unbinned signal-to-noise ratios for the BAO (middle panel) and for the whole spectra (lower panel). Three different choices for the size of the photo-z bin are shown. The lines with three different thickness levels, starting from the thickest, assume the optical follow-up parameters (i) $\sigma_0=10^{-3}$, $f_{\rm fail}=0.0$, (ii) $\sigma_0=10^{-2}$, $f_{\rm fail}=0.0$, and (iii) $\sigma_0=10^{-2}$, $f_{\rm fail}=0.3$, respectively. In the upper panel for the first of the above three cases binned $1\sigma$ error boxes are also shown. The light gray vertical stripe marks the small-scale modes excluded from our analysis.}
\label{fig6}
\end{figure}

\subsection{Example spectra}
In the upper panel of Fig.~\ref{fig6} we show angular auto-spectra for three different choices for the width of the photo-z bin as specified in the upper righthand corner. Here we have assumed a full sky coverage down to a soft-band limiting flux of $10^{-14}$ erg/s/cm$^2$. Thick, middle-thick, and thin lines correspond to the optical follow-up with (i) $\sigma_0=10^{-3}$, $f_{\rm fail}=0.0$, (ii) $\sigma_0=10^{-2}$, $f_{\rm fail}=0.0$, and (iii) $\sigma_0=10^{-2}$, $f_{\rm fail}=0.3$, respectively. Only for the first of the above three cases do we show binned $1\sigma$ error boxes. The small-scale modes with $\ell > \ell_{\max}=500$, which are not included in our analysis, are marked with a gray shaded band.

\begin{figure*}
\centering
\includegraphics[width=\plotwdtwo]{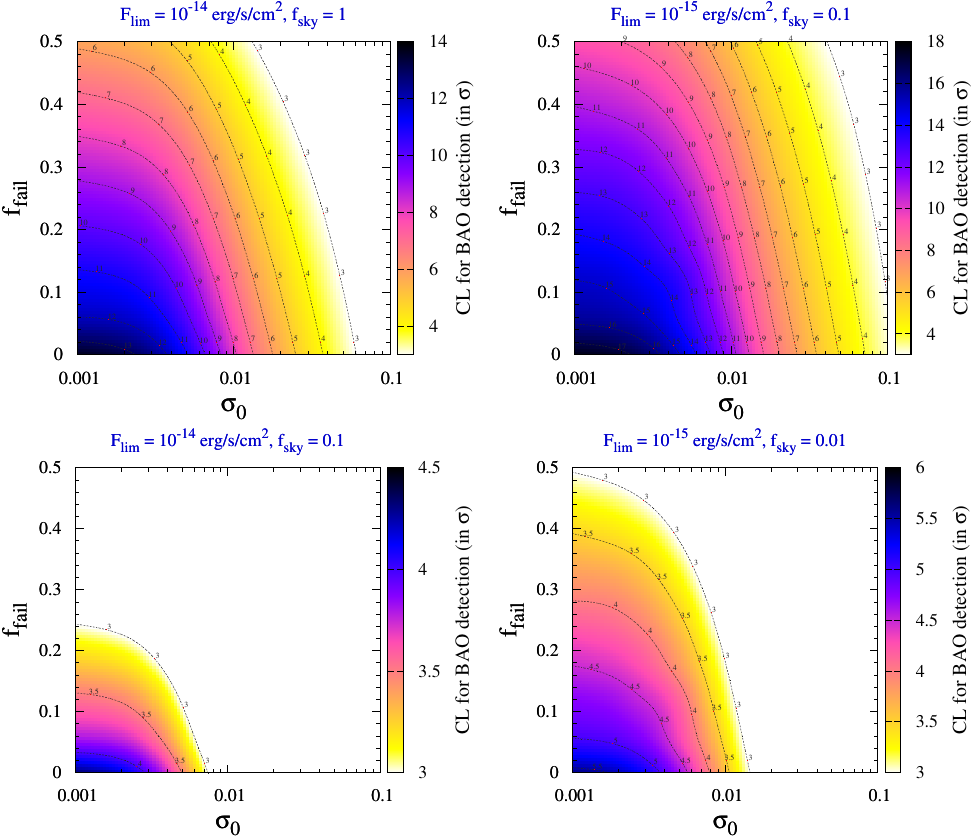}
\caption{Confidence levels for the BAO detection as a function of the quality of the optical follow-up, specified by $\sigma_0$ and $f_{\rm fail}$, for limiting soft-band X-ray fluxes $F_{\rm lim}=10^{-14}$ (lefthand panels) and $10^{-15}$ erg/s/cm$^2$ (righthand panels). For $F_{\rm lim}=10^{-14}$ ($10^{-15}$) cases with a complete and $10\%$ ($10\%$ and $1\%$) sky coverage are shown. A redshift range $z=0.5-2.5$ is assumed.}
\label{fig7}
\end{figure*}

\begin{figure*}
\centering
\includegraphics[width=16.5cm]{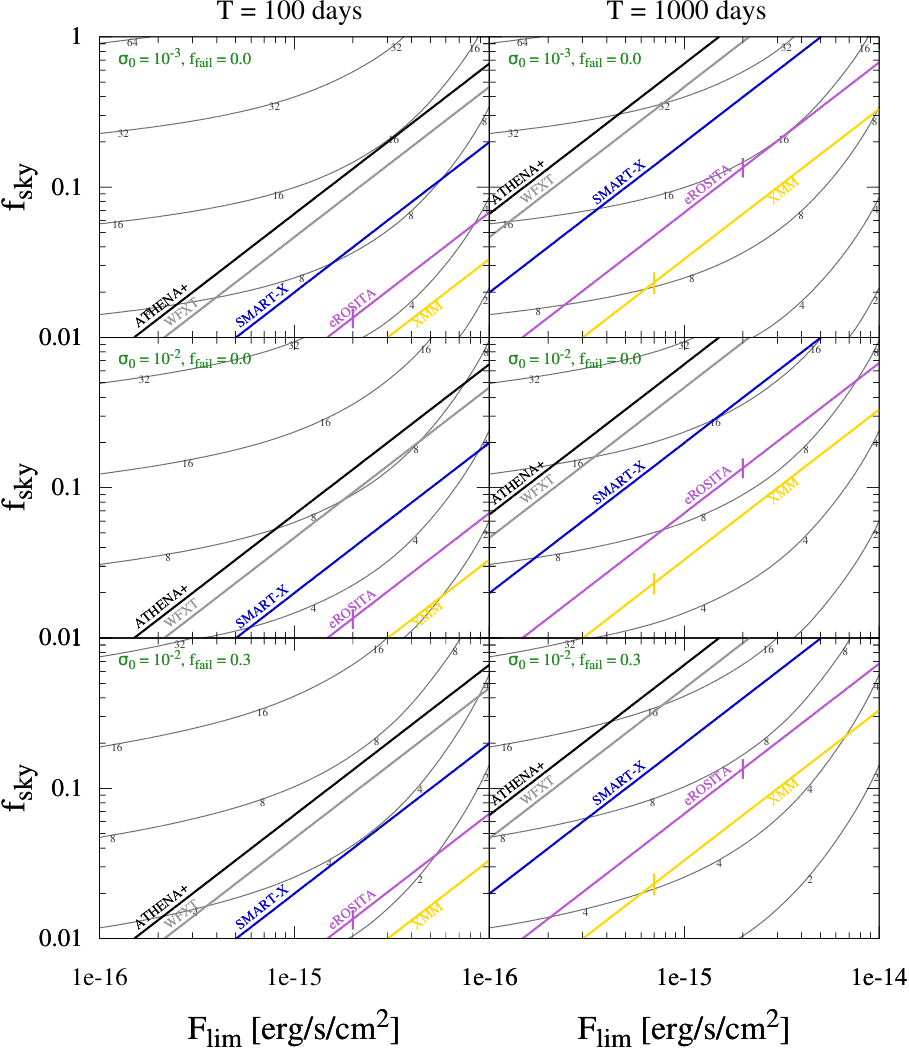}
\caption{Achievable signal-to-noise for the BAO detection as a function of limiting flux and sky coverage, shown with numbered contour lines. From top to bottom, three different choices for the quality of optical follow-up are considered. The diagonal solid lines show possible survey size-depth relations for several existing and proposed X-ray instruments assuming a total observation time of $100$ days (lefthand panels) and $1000$ days (righthand panels). Redshift range $z=0.5-2.5$ is assumed. For eROSITA and XMM-Newton, the small vertical lines at $F_{\rm lim}\simeq 2\times 10^{-15}$ and $7\times 10^{-16}$ erg/s/cm$^2$ mark approximate confusion noise limits for these instruments. For the other telescopes, confusion noise is not achieved in the considered flux range.}
\label{fig8}
\end{figure*}

One can see how the angular clustering strength increases after reducing the size of the photo-z bin, since the radial selection function gets correspondingly more sharply peaked. This behavior is qualitatively understandable, because a smaller radial projection length leads to less smoothing and thus keeps a higher level of coherence in the projected field. More quantitatively, from Eq.~(\ref{eq10}) one obtains -- for a narrow selection function (i.e., can be replaced with a delta function) and for sufficiently high $\ell$, so that the spherical Bessel function can be assumed to be sharply peaked at $\ell=kr$ (also for high $\ell$ the redshift-distortion contribution vanishes) -- $C_{\ell}\propto \frac{P}{r\Delta r}$, with $r$ and $\Delta r$ the comoving distance and its interval, corresponding to $z$ and $\Delta z$, respectively. Effective $\Delta r$ is increased by choosing a wider photo-z bin and also by having higher values for $\sigma_0$ and $f_{\rm fail}$, i.e., poorer quality photo-z.

Even though the signal has the highest amplitude for the narrowest photo-z bin shown in Fig.~\ref{fig6}, the noise also starts to increase considerably owing to the lower number of AGN available in narrow redshift bin. This is more visible in the middle and lower panels of Fig.~\ref{fig6} where we show the unbinned signal-to-noise for the BAO part and for the whole spectrum, respectively. One can see that the signal-to-noise for the full spectrum is approximately an order of magnitude higher than for the BAO alone, since the BAO represents only a $\sim 5\%$ fluctuation on top of the smooth broad-band spectrum.

Although an increased Poisson noise leads to a reduced signal-to-noise per redshift bin in the case of narrow redshift bins, by joining the information from the larger total number of available bins, one is able to achieve generally higher cumulative signal-to-noise, as is shown in the next section.

\section{Main results}\label{sec4}

The main results of this paper are shown in Figs.~\ref{fig7} and \ref{fig8}. In Fig.~\ref{fig7} the limiting soft-band flux and survey area, i.e. the parameters describing X-ray survey, are kept fixed, while the parameters describing optical follow-up $\sigma_0$ and $f_{\rm fail}$ are allowed to vary within ranges $10^{-3}-10^{-1}$ and $0.0-0.5$, respectively. The range of redshifts is constrained to be between $0.5$ and $2.5$. In Fig.~\ref{fig8} the situation is reversed: the follow-up parameters are kept fixed and X-ray survey parameters, $F_{\rm lim}$ and $f_{\rm sky}$, vary.

In Fig.~\ref{fig7} we have considered two values for $F_{\rm lim}$: $10^{-14}$ and $10^{-15}$ erg/s/cm$^2$. The first of these values is achievable by the eRASS for the entire extragalactic sky,  while the second value is more characteristic of the deeper surveys, such as XMM-COSMOS; a comparable sensitivity may be also achieved in the polar regions of eRASS covering about $\sim 100$ deg$^2$. For $F_{\rm lim}=10^{-14}$ and $10^{-15}$, we have presented cases with $f_{\rm sky}=\{1,0.1\}$ and $\{0.1,0.01\}$, respectively.

For example, for a survey with $F_{\rm lim}=10^{-14}$ and full sky coverage, $\gtrsim 5\sigma$ detection of the BAO is achievable once $\sigma_0\sim 0.01$ and $f_{\rm fail}\lesssim 0.3$. A similar quality optical follow-up in combination with $F_{\rm lim}=10^{-15}$ and $f_{\rm sky}=0.1$ would lead to $\gtrsim 8\sigma$ detection. With $F_{\rm lim}=10^{-14}$ and $10\%$ sky coverage, the BAO can only be detected provided very high-accuracy photometric redshifts or a full spectroscopy is available. If $1\%$ of the sky is covered down to $F_{\rm lim}=10^{-15}$, photometric redshifts with $\sigma_0\sim 0.01$ and $f_{\rm fail}\sim 0$ should lead to a marginal BAO detection with $\sim 3\sigma$.

In Fig.~\ref{fig8} we investigate the dependence of the CL for the BAO detection as a function of the survey parameters. The contour lines labeled with numbers show achievable CL  as a function of $F_{\rm lim}$ and $f_{\rm sky}$, while keeping the follow-up parameters $\sigma_0$ and $f_{\rm fail}$ fixed to the values as denoted at the top lefthand corners of the panels. In addition to the above contour lines, which are the same for respective left- and righthand panels, we show the loci of possible surveys with the XMM-Newton\footnote{\url{http://xmm.esac.esa.int/}} (the only existing X-ray instrument with reasonable survey capabilities) and several upcoming/proposed X-ray instruments, assuming total survey duration of $100$ days (lefthand panels) and $1000$ days (righthand panels).

These  lines are calculated as follows. The fraction of sky covered by a survey can be expressed as
\begin{equation}
f_{\rm sky}=\frac{T}{t}f_{\rm FoV}\,,
\label{eq:fsky}
\end{equation}
where $T$ is the total survey duration, $t$ the time  spent per pointing, and $f_{\rm FoV}$ the fraction of sky covered by the instrument's field of view (FoV). We assume that limiting flux for the AGN detection scales as
\begin{equation}
F_{\rm lim}\propto (A_{\rm eff}\times t)^{-1}\,,
\label{eq:flim}
\end{equation}
where $A_{\rm eff}$ is the soft-band effective area of the instrument.
In reality, limiting flux for a point source  detection also depends on the angular resolution of the instrument (via background)  and parameters of its orbit around the Earth. For the purpose of this illustrative calculation, we ignore these complications.

Since for most of the future X-ray instruments considered in this work, in particular for SMART-X, WFXT, and ATHENA+, the angular resolution is sufficiently high, confusion is not a problem for the fluxes considered in this study. However,  having similar angular resolution of $\sim 15^{''}$ `on axis', observations with eROSITA and XMM-Newton, become confusion-noise-dominated at fluxes below $F_{\rm lim}\sim 1-2\times 10^{-15}$ erg/s/cm$^2$. In Fig.~\ref{fig8} the approximate confusion noise limits for eROSITA and XMM-Newton are marked by small vertical lines on the corresponding curves.

With equations (\ref{eq:fsky}) and (\ref{eq:flim}), we can write
\begin{equation}\label{eq25}
f_{\rm sky}\propto T F_{\rm lim}A_{\rm eff}f_{\rm FoV}\,.
\end{equation}
Here the combination $A_{\rm eff}f_{\rm FoV}$ is the grasp of the instrument. The above scaling is normalized such that for eRASS, with $T=4$ yr and with the grasp taken from the instrument's webpage \url{http://www.mpe.mpg.de/erosita/}, the limiting flux reults in $F_{\lim}=1.1\times 10^{-14}$ erg/s/cm$^2$ for $f_{\rm sky}=1$ \citetext{see, e.g., \citealp{2013A&A...558A..89K}}. The values of grasp in the soft band for ATHENA+\footnote{\url{http://the-athena-x-ray-observatory.eu/}}, WFXT\footnote{\url{http://wfxt.pha.jhu.edu/}}, SMART-X\footnote{\url{http://smart-x.cfa.harvard.edu/}}, and XMM-Newton were computed from the parameters given in the instruments' manuals and websites. The particular numbers for $A_{\rm eff}$ and FoV used in our calculations are given in Table~\ref{tab1}.

Only for eROSITA have we used $A_{\rm eff}$ appropriately averaged over the instrument's FoV and over the energy range of $0.5-2$ keV, with the additional assumption that a typical AGN has a photon index $\Gamma=1.9$\footnote{Thus, the effective area is calculated as \[A_{\rm eff}=\frac{\int{\rm d}E\, E^{-\Gamma}\int_{{\rm FoV}}{\rm d}\Omega\, A_{\rm eff}(E,\hat{\bf{n}})}{\int{\rm d}E\, E^{-\Gamma}\int_{{\rm FoV}}{\rm d}\Omega}\,,\] where $\hat{\bf{n}}$ is a unit vector specifying position inside the FoV, and ${\rm d}\Omega$ is a solid angle element.} The value of $A_{\rm eff}$ calculated this way for eROSITA is used to obtain appropriate normalization factors in Eqs.~(\ref{eq25}) and (\ref{neweq02}). For the other instruments, the values of $A_{\rm eff}$ as given in Table~\ref{tab1} are slightly overestimated, since averaging over the FoV would certainly reduce them somewhat. Since for the future instruments, we have no information available regarding the behavior of the `off axis' effective area, we used the `on axis' $A_{\rm eff}$ at $E\sim 1$ keV, for simplicity in these cases.

\begin{table}
\centering
\caption{Assumed values of the effective area and size of the field of view for the X-ray instruments of Fig.~\ref{fig8}.}
\label{tab1}
\begin{tabular}{l|l|l|}
 & {\scriptsize $A_{\rm eff}$ $[{\rm cm}^2]$} & {\scriptsize FoV $[{\rm deg}^2]$} \\
\hline
{\scriptsize XMM-Newton} & $2000$ & $0.25$\\
{\scriptsize eROSITA} & $1200\,^{(*)}$ & $0.85$\\
{\scriptsize SMART-X} & $20000$ & $0.15$\\
{\scriptsize WFXT} & $7000$ & $1.0$\\
{\scriptsize ATHENA+} & $20000$ & $0.5$\\
\hline
\end{tabular}\\\hspace{-3.cm}
${\tiny ^{(*)}}$ {\scriptsize FoV-averaged}
\end{table}

From Fig.\ref{fig8} one can see that for eROSITA type of instrument an all-sky survey would not be the best strategy to study BAO. In particular, for a $T=1000$ day survey, the optimal BAO detection of $\sim 16\sigma$ will be achieved when covering $f_{\rm sky}\sim 0.2$ of the sky, i.e. the sky area of $\sim 9000$ deg$^2$. The optimal sky area does not strongly depend on the parameters of the optical followup within the considered range, but the BAO detection significance does. For the actual eRASS parameters ($T=4$ yr $\sim 1500$ days, $f_{\rm sky}\sim 1$),   BAO should be detected at the significance levels of  $\sim 14\sigma$, $\sim 8\sigma$, and $\sim 5\sigma$, for the three sets of the optical followup parameters considered in the upper, middle, and bottom panels in Fig. \ref{fig8}, respectively. These numbers are somewhat larger than the ones found in \citet{2013A&A...558A..90K}, mostly due to including the information carried by the cross-spectra. In addition to cross-spectra the other differences  compared to \citet{2013A&A...558A..90K} analysis were described at the beginning of section~\ref{sec3}.

From Fig.~\ref{fig8} we can also see that for the next generation of X-ray telescopes, such as ATHENA+ or WFXT, only a $100$-day survey covering $\sim 10-20\%$ of the sky could lead to a $6\sigma-13\sigma$ BAO detection, assuming the optical follow-up with quality parameters shown there. Also, a $T=100$ day survey with XMM over less than $\sim 1\%$ of the sky, plus a high-accuracy photometric or full spectroscopic follow-up, could lead to a marginal $\sim 3\sigma$ detection, which in reality is quite comparable to most of the existing BAO measurements in the optical band, but probes  the BAO at $z\sim 1$, which has not been covered by any survey thus far.

It is interesting to estimate the cost of these surveys in terms of the optical follow-up, i.e. in terms of the number of objects requiring accurate redshift determination. For an optimal eROSITA survey with duration $T=1000$ days, the sensitivity of $\sim 3\times 10^{-15}$ erg/s/cm$^2$ needs to be reached over an area of $\sim 9000$ deg$^2$. At this flux limit, the AGN density is $\sim 350$ deg$^{-2}$, giving in total  about $\sim 3\times 10^{6}$ objects for a follow-up. This number is comparable to the total number of AGN to be detected in the planned $T=4$ yr all-sky survey by eROSITA, but is located in a roughly five times smaller sky area, which will simplify the optical follow-up. A similar number of objects for the optical follow-up over similar sky area  will be produced  by a ``BAO-optimized'' survey by ATHENA+ with duration $T=100$ days. On the other hand, a $T=100$ day BAO survey with XMM-Newton down to the same optimal limiting flux will cover $\sim 440$ deg$^{2}$ and provide $\sim 1.5\times 10^5$ objects in total for the optical follow-up. This is an entirely feasible task, given the progress in the instrumentation for multiobject spectroscopy. The optical follow-up requirements for a ATHENA+ survey (about $\sim 3\times 10^6$ objects over $\sim 9000$ deg$^2$) also do not seem to be entirely beyond reach, since they are similar in the sky area and exceed the recent  BOSS survey by  a factor of just a few in the object density.

\subsection*{Optimal survey strategy}
In the following we provide the results for the optimal survey strategy in a compact form. 

In Fig.~\ref{fig8} the the optimal surveys correspond to the points where the instrumental lines are tangent to the signal-to-noise curves. As one can see from this figure, up to a good approximation, the optimal limiting flux turns out not to depend on any particular instrument used to perform observations. It also weakly depends on the parameters of the optical follow-up, within the considered range. Once the optimal limiting flux is reached, instead of integrating at the same pointing direction any longer, one should move to the new field.

For the follow-up parameters $\sigma_0=10^{-3}$ \& $f_{\rm fail}=0.0$, the optimal limiting flux is
\begin{equation}\label{neweq01}
F_{\rm lim}^{\rm optim}\simeq 3.2\times 10^{-15}\, {\rm erg/s/cm^2}\,.
\end{equation}
Having fixed $F_{\rm lim}$ to the above value, the survey area can now be expressed as
\begin{equation}\label{neweq02}
A[{\rm deg}^2]\simeq 220 \left[\frac{T}{1000\,{\rm days}}\right]\left[\frac{{\rm FoV}}{0.25\,{\rm deg}^2}\right]\left[\frac{A_{\rm eff}}{1000\,{\rm cm}^2}\right]\,,
\end{equation}
and the CL for the BAO detection as
\begin{equation}\label{neweq03}
{\rm CL} \simeq 5.4\sqrt{\frac{A}{1000\,{\rm deg}^2}}\quad{\rm sigma}.
\end{equation}
The extragalactic source density at $F_{\rm lim}\simeq 3.2\times 10^{-15}$ is $\simeq 350$ deg$^{-2}$; i.e., the number of sources above this limiting flux can be estimated as
\begin{equation}\label{neweq04}
N \simeq 350\times A[{\rm deg}^2]\,.
\end{equation}
Here we assume soft-band number counts taken from \citet{Kim2007}.

For the other two sets of follow-up parameters discussed in this paper; i.e., $\sigma_0=10^{-2}$ and $f_{\rm fail}=0.0$, and $\sigma_0=10^{-2}$ and $f_{\rm fail}=0.3$, the optimal limiting fluxes are $F_{\rm lim}\simeq 2.9\times 10^{-15}$ and $2.4\times 10^{-15}$ erg/s/cm$^2$, respectively. In the first case, this leads to the replacements $220\rightarrow 200$, $5.4\rightarrow 3.5$, and $350\rightarrow 380$ in Eqs.~(\ref{neweq02}), (\ref{neweq03}), and (\ref{neweq04}), respectively. For $\sigma_0=10^{-2}$ and $f_{\rm fail}=0.3$, the corresponding replacements are $220\rightarrow 170$, $5.4\rightarrow 2.8$, $350\rightarrow 440$.

\begin{figure}
\centering
\includegraphics[width=\plotwd]{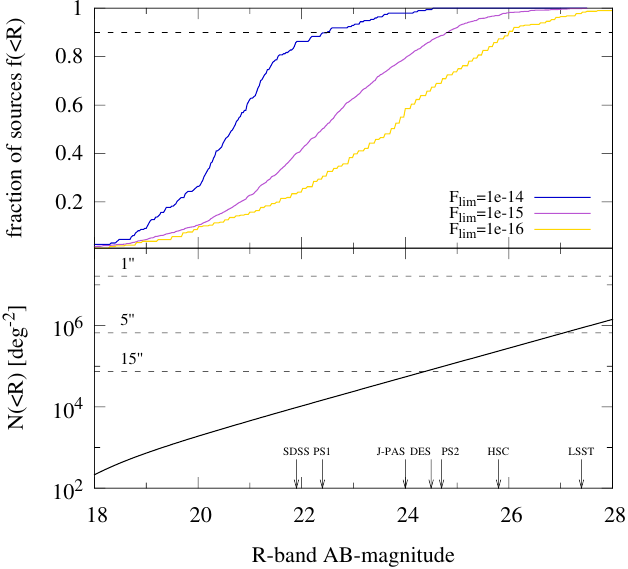}
\caption{\emph{Upper panel:} cumulative R-band magnitude distribution of optical counterparts of X-ray sources in XMM-COSMOS and CDF-N fields for the X-ray limiting fluxes of $10^{-14}$, $10^{-15}$, and $10^{-16}$ erg/s/cm$^2$. \emph{Lower panel:} cumulative number density of galaxies as a function of their R-band magnitude. The horizontal dashed lines show the number of `resolution elements' per deg$^2$ corresponding to angular error circle diameters of $1''$, $5''$, and $15''$. The approximate limiting magnitudes for various existing and future optical imaging surveys are also displayed.}
\label{fig9}
\end{figure}

\subsection*{Optical identification}
In the upper panel of Fig.~\ref{fig9} we show the cumulative R-band AB-magnitude distributions for X-ray AGN brighter than the limiting fluxes as specified in the legend. For the flux limits of $10^{-14}$ and $10^{-15}$, we used the optical follow-up data of the XMM-COSMOS field as provided by \citet{2010ApJ...716..348B}. For $F_{\rm lim}=10^{-16}$ the Chandra Deep Field North (CDF-N) data from \citet{2008ApJS..179....1T} is used. For the CDF-N the follow-up data includes the R-band AB-magnitudes, while for XMM-COSMOS, the corresponding R-band magnitudes are estimated using available $r$ and $i$-band data and applying a photometric system conversion relation by Lupton (2005) as given in \url{http://classic.sdss.org/dr7/algorithms/sdssUBVRITransform.html}.

One can see that for the limiting fluxes of $10^{-14}$, $10^{-15}$, and $10^{-16}$ erg/s/cm$^2$, $90\%$ of the optical counterparts are brighter than $m_{\rm R}\simeq 22.5$, $\sim 25$, and $\sim 26$, respectively. The approximate magnitude limits for several existing and upcoming optical imaging surveys are given in the lower panel of Fig.~\ref{fig9}.\footnote{The magnitude limits were converted from the $r$ to a somewhat broader R-band by simply assuming an effective spectral index of $\alpha=1$, i.e. $F_\lambda\propto \lambda^{-\alpha}$ at $\lambda\sim 600$ nm, and taking differences in filter widths and their centroid locations into account. This leads to an approximate relation ${\rm R}\simeq r-0.3$.} Thus, for the X-ray surveys with $F_{\rm lim}=10^{-14}$ erg/s/cm$^2$, e.g. for the eRASS, $\sim 80\%$, and $\sim 90\%$ of the optical counterparts should already be available inside the existing SDSS and Pan-STARRS\footnote{\url{http://pan-starrs.ifa.hawaii.edu}} (PS1) survey footprints, while for surveys with $F_{\rm lim}=10^{-15}$ or $10^{-16}$, deeper photometry (e.g., J-PAS\footnote{\url{http://j-pas.org/}}, DES\footnote{\url{http://www.darkenergysurvey.org/}}, PS2, HSC\footnote{\url{http://www.naoj.org/Projects/HSC/}}, LSST\footnote{\url{http://www.lsst.org/}}) is needed.

Depending on the angular resolution of the X-ray instrument, there might be serious difficulties correctly identifying the optical counterpart in the imaging data. In the lower panel of Fig.~\ref{fig9} we have plotted the cumulative number counts of galaxies in the R band. This model curve is obtained as a least squares fit to the galaxy number-count data \citep{1991MNRAS.249..481J,1991MNRAS.249..498M,1995MNRAS.273..257M,2000MNRAS.311..707M,2001MNRAS.323..795M} compiled by Nigel Metcalfe, which is accessible at \url{http://astro.dur.ac.uk/~nm/pubhtml/counts/counts.html}. With horizontal dashed lines we show the number of `resolution elements' per square degree corresponding to angular error circle diameters of $1''$, $5''$, and $15''$. As long as these lines stay above the solid number count curve, on average there is fewer than one object per resolution element, and thus the source identification should be possible. For the planned future instruments such as ATHENA+, WFXT, or SMART-X, point sources should be easily detectable with angular accuracy better than $1''$, and so there should be no significant confusion for finding optical counterparts for X-ray surveys with limiting fluxes as low as $10^{-16}$ erg/s/cm$^2$. 
Even for eROSITA, with relatively broad PSF, the positional accuracy for point sources can be achieved at the level of better than $10''$, possibly up to $\sim 5''$ (A.~Merloni, private communication). Thus, for the limiting flux of $F_{\rm lim}=10^{-14}$, one should expect one (optical) galaxy per $\gtrsim 10$ resolution elements. Therefore majority of the optical counterparts should be unambiguously identifiable.

\subsection*{Optical follow-up}
In comparison to the optical imaging surveys as listed in Fig.~\ref{fig9}, the existing wide field spectroscopic surveys are significantly shallower; e.g., the BOSS QSO sample reaches ${\rm R}\lesssim 21.5$. However, the planned wide field survey by DESI\footnote{\url{http://desi.lbl.gov/}} \citep{2013arXiv1308.0847L} should get spectra for QSOs as faint as ${\rm R}\lesssim 23$. Thus, already with a $4m$ class telescope, it is possible to efficiently follow up X-ray-selected AGN samples with $F_{\rm lim}=10^{-14}$. In principle, DESI could easily follow up approximately half of the total number of X-ray AGN detected in eRASS by allocating $\sim 15\%$ of the fibers to these sources and staying within the planned sky coverage of $\sim 40\%$. A follow-up program on somewhat smaller scale, called SPIDERS\footnote{\url{http://www.sdss3.org/future/spiders.php}}, is already planned within an upcoming SDSS-IV project.

The other very exciting possibility for following up on eRASS AGN is with narrow band photometric surveys like J-PAS \citep{2014arXiv1403.5237B}. J-PAS with its $54$ narrow and $2$ broad-band filters should reach an effective depth of ${\rm R}\simeq 24$, and in narrow-band filters it goes down to $\sim 22-23$ magnitude. However, the achievable photo-z accuracy for the X-ray-selected AGN still needs to be investigated in full detail.

For the optical follow-up of X-ray AGN samples with $F_{\rm lim}\sim 10^{-16}-10^{-15}$, it should be clear that one needs to use $6-10$m class telescopes for efficient spectroscopic follow-up.

\subsection*{Dependence on the assumed effective halo mass $M_{\rm eff}$}
Throughout this study we have used $M_{\rm eff}=2\times 10^{13}$ M$_\odot$ as our default value, which as stated above, is fully compatible with the clustering measurements of the X-ray-selected AGN. However, to get a feeling for how much our results could change, once different values for $M_{\rm eff}$ are assumed, we give values for the signal-to-noise ratios corresponding to the lower lefthand corners of the panels in Fig.~\ref{fig7}, i.e., for the cases with low resolution spectroscopy or high accuracy photo-z with negligible catastrophic error fraction. For our standard assumption of $M_{\rm eff}=2\times 10^{13}$ M$_\odot$, the confidence levels for the BAO detection are $\sim 14\sigma$, $\sim 18\sigma$, $\sim 5.6\sigma$, and $\sim 4.3\sigma$, starting clockwise from the upper leftmost panel. In case one adopts $M_{\rm eff}=5\times 10^{12}$ M$_\odot$, the corresponding numbers reduce to $\sim 8.1\sigma$, $\sim 13\sigma$, $\sim 4.1\sigma$, and $\sim 2.6\sigma$. If $M_{\rm eff}=10^{13}$ M$_\odot$ ($5\times 10^{13}$ M$_\odot$), one obtains $\sim 10\sigma$ ($\sim 20\sigma$), $\sim 15\sigma$ ($\sim 20\sigma$), $\sim 4.8\sigma$ ($\sim 6.5\sigma$), and $\sim 3.3\sigma$ ($\sim 6.2\sigma$).

\section{Summary and conclusions}\label{sec5}

We investigated the potential of large samples of X-ray-selected AGN to probe the large-scale structure of the Universe, in particular, to detect the BAO.  For the X-ray-selected AGN, most of the BAO signal comes from redshifts $z\sim 1$, where the X-ray AGN population peaks. These redshifts are currently uncovered by any one of the existing dedicated BAO surveys.
However,  although X-ray surveys are very efficient in producing large samples of AGN,  they  do not provide any redshift information and so have to be complemented with an optical follow-up. 

The main goals of this study were (i) to find out the required quality criteria for the optical follow-up and (ii) to formulate the optimal strategy of the X-ray survey  in order to facilitate accurate measurements of the clustering two-point function and detection of the BAO.

Our main results are presented in Figs.~\ref{fig7} and \ref{fig8}, where the confidence levels for the BAO detection are shown as a function of X-ray survey parameters ($F_{\rm lim}$, $f_{\rm sky}$) and the parameters describing the quality of the optical follow-up ($\sigma_0$, $f_{\rm fail}$).
In particular we demonstrated that redshift accuracy of $\sigma_0=10^{-2}$ at $z=1$ and the failure rate of $f_{\rm fail}\la 30\%$ are sufficient for a reliable  detection of  BAO in the future X-ray surveys. If spectroscopic quality redshifts ($\sigma_0=10^{-3}$ and  $f_{\rm fail}\sim 0$) are available, the confidence level of the BAO detection will be boosted by a factor of $\sim 2$.  

For the meaningful detection of BAO, X-ray surveys   of moderate depth of $F_{\rm lim}\sim {\rm a\,\,few~} 10^{-15}$ erg/s/cm$^2$ covering sky area from a few hundred to a few ten thousand square degrees are required. For the fixed survey duration, the optimal strategy for the BAO detection does not necessarily require full sky coverage. For example, in a $T=1000$ day survey by an eROSITA type telescope, an optimal strategy for BAO detection requires a survey of $\sim 9000$ deg$^2$ and would yield a $\sim 16\sigma$ BAO detection.   A similar detection will be achieved by ATHENA+ or WFXT type telescopes in a survey with a duration of 100 days, covering similar sky area. XMM-Newton can achieve a marginal BAO detection in a 100-day survey covering $\sim 400$ deg$^2$. 

These surveys would impose moderate to high demands on the optical follow-ups requiring determination of redshifts of $\sim 10^5$ objects (XMM-Newton) to $\sim 3\times 10^6$ objects (eROSITA, ATHENA+ and WFXT) in the above-mentioned sky areas. Given the progress in the instrumentation for multi-object spectroscopy, these demands appear to be within the reach of modern and future ground-based optical facilities.

Since the BAO is $\sim 5\%$ modulation on top of a smooth broad-band spectral shape, the amplitude of the power spectrum, hence the AGN clustering bias, can be determined with an order-of-magnitude higher signal-to-noise, and so can be done with much poorer quality photometric redshifts.

\begin{appendix}
\section{`No wiggle' spectral template}\label{appa}

\begin{figure}
\centering
\includegraphics[width=\plotwd]{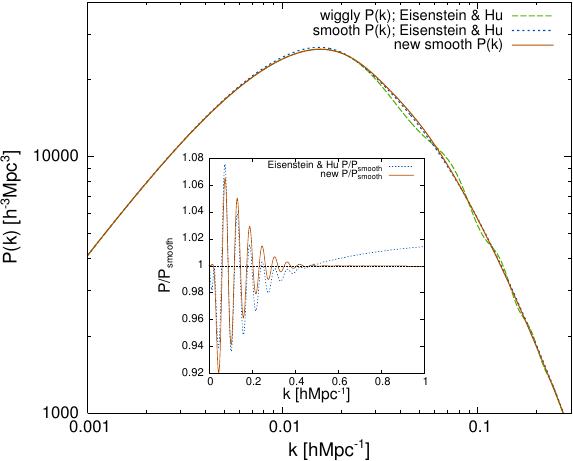}
\caption{Smooth spectral template used in this paper. The blue dashed line shows a similar smooth template proposed by \citet{1998ApJ...496..605E}}
\label{fig10}
\end{figure}

To isolate the BAO signal, we do not use the usual `no wiggle' spectral form of \citet{1998ApJ...496..605E}. Instead, we divide the power spectrum with a smooth model that is calculated essentially the same way as the `wiggly' case, with only the following replacement in Eq.~(21) of \citet{1998ApJ...496..605E}
\begin{equation}
j_0(x)=\frac{\sin(x)}{x}\longrightarrow \exp\left(-\frac{x^2}{6}\right)\,.
\end{equation}
With this replacement the peak region of $j_0(x)$ at small $x$ is matched well (up to ${\cal O}(x^4)$) and for larger arguments, the exponent efficiently smoothes out the oscillatory features. 

The resulting smooth spectrum and extracted BAO are shown in Fig.~\ref{fig10} in comparison to the original \citet{1998ApJ...496..605E} form. Since in this study we apply an effective wavenumber cutoff of $k_{\max}\sim 0.1$ $h\,$Mpc$^{-1}$, both smooth spectral forms lead to almost indistinguishable results.

\end{appendix}

\acknowledgements{We thank our referee for suggestions that greatly helped to improve the paper. GH acknowledges support by the Estonian Ministry of Education and Research grant IUT26-2. MG acknowledges hospitality of the Kazan Federal University (KFU) and support by the Russian Government Program of Competitive Growth of KFU.}
\bibliographystyle{aa}
\bibliography{references}

\end{document}